\documentclass[%
 aip,
 jcp,
 amsmath,amssymb,
 reprint,%
longbibliography 
]{revtex4-1}
\usepackage[%
                  bookmarks,%
                  bookmarksopen=false,
                  pdfauthor={Autor},%
                  pdftitle={Titel},%
                  colorlinks=true,
                  linkcolor=blue,
                  citecolor=red,%
                  urlcolor=cyan,%
                ]{hyperref}
\usepackage{refcount}
\usepackage{graphicx}
\usepackage{dcolumn}
\usepackage{bm}
\usepackage{braket}

\usepackage[T1]{fontenc}
\usepackage[utf8]{inputenc}
\usepackage{mathptmx}
\draft 

\usepackage{caption}
\usepackage{subcaption}
\newcommand*{\B}{~\text{B}\ensuremath{_0}}
\usepackage{siunitx}
\usepackage{booktabs}
\usepackage{multirow, makecell}
\usepackage{graphicx}
\newcommand{\abs}[1]{\lvert#1\rvert}
\usepackage{chemfig}
\usepackage{chemexec}
\begin{document}

\preprint{AIP/123-QED}

\title[]{Cholesky decomposition of complex two-electron integrals over GIAOs: Efficient MP2 computations for large molecules in strong magnetic fields}

\author{Simon Blaschke}
 \email{siblasch@uni-mainz.de}
 
\author{Stella Stopkowicz}%
 \email{sstopkow@uni-mainz.de}
 \affiliation{Department Chemie, Johannes Gutenberg-Unversität Mainz, Duesbergweg 10-14, D-55128 Mainz, Germany}%

\date{\today}

\begin{abstract}
In large-scale quantum-chemical calculations the electron-repulsion integral (ERI) tensor rapidly becomes the bottleneck in terms of memory and disk space. 
When an external finite magnetic field is employed, this problem becomes even more pronounced because of the reduced permutational symmetry and the need to work with complex integrals and wave-function parameters. 
One way to alleviate the problem is to employ a Cholesky decomposition (CD) to the complex ERIs over gauge-including atomic orbitals. The CD scheme establishes favourable compression rates by selectively discarding linearly dependent product densities from the chosen basis set while maintaining a rigorous and robust error control. 
This error control constitutes the main advantage over conceptually similar methods such as density fitting which rely on employing pre-defined auxiliary basis sets. 
We implemented the use of the CD in the framework of finite-field (ff) Hartree-Fock and ff second-order M\o ller Plesset perturbation theory. Our work demonstrates that the CD compression rates are particularly beneficial in calculations in the presence of a finite magnetic field. The ff-CD-MP2 scheme enables the correlated treatment of systems with more than 2000 basis functions in strong magnetic fields within a reasonable time span.
\end{abstract}

\maketitle

\section{Introduction}\label{sec:introduction}
When going towards larger systems in quantum-chemical ab-initio calculations, the standard formulations of the respective methods quickly become computationally too expensive. 
One of the most common bottlenecks is the evaluation and handling of the electron-repulsion integrals (ERIs) over one-electron basis functions due to their formal quartic scaling with the number of atomic orbitals. 
Therefore, various methods have been developed to approximate the two-electron integral tensor or make its calculation  more efficient.\\
For example, integral prescreening\cite{Ahlrichs.J.Comp.Chem.1989,Ochsenfeld.J.Chem.Phys.2005,Ochsenfeld.J.Chem.Phys.2005erratum} methods drastically reduce the number of relevant integrals by discarding elements of the ERI tensor which are smaller than a chosen threshold. This is achieved without their explicit evaluation by utilising Schwarz's inequality. For large systems, the number of significant integrals then scales quadratically with the number of basis functions.\cite{DyczmonsTheoret.Chim.Acta1973,MolecularElectronicStructure}\\
The computational effort can be reduced further by linear-scaling techniques such as the fast multipole moment\cite{Greengard.1987,HeadGordon.1994} (FMM) method which exploits the classical electrostatic character of the electron repulsion for well separated charges and enables the description of the Coulomb interaction with linear scaling.
However, linear-scaling techniques (and prescreening) solely lead to a reduction in computational cost for very large systems with more than several thousand basis functions, since the scaling behaviour is dominated by a large prefactor.\cite{NeeseOnLinearScaling}\\
A further approach which shows a beneficial scaling behaviour also for small and medium-sized molecules is density fitting, also called resolution-of-identity approximation (RI).\cite{Whitten.1973,Almloef.Chem.Phys.Lett.1993,Feyereisen.Chem.Phys.Lett.1993,Aquilante.Chem.Phys.Lett.2007,Reine.J.Chem.Phys.2008,Aquilante.J.Chem.Phys.2009,Pedersen.Theor.Chem.Acc.2009} Additionally, any tensor such as the full ERI or likewise the associated RI representation can be further factorised through the use of tensor hypercontractions.\cite{Martinez.J.Chem.Phys.I.2012,MartinezJ.Chem.Phys.II.2012,MartinezJ.Chem.Phys.III.2012,MartinezJ.Chem.Phys.2014,OchselfeldJ.Chem.TheoryComput.2020}
The performance of the RI approach is tied to the quality of an externally optimized auxiliary basis set. A more general approach which does not require an auxiliary basis, is the Cholesky decomposition (CD).\cite{LinderbergJ.Quantum.Chem.1977,KochJ.Chem.Phys.2003,KochJ.Chem.Phys.2019} 
The use of CD in quantum chemistry was first suggested by Beebe \& Linderberg\cite{LinderbergJ.Quantum.Chem.1977} and used several years later together with an efficient implementation of a partial pivoting Cholesky algorithm by Koch et al.\cite{KochJ.Chem.Phys.2003}
Recently, a yet more sophisticated decomposition algorithm has been proposed.\cite{KochJ.Chem.Phys.2019}
After initial use in ground-state methods such as Hartree-Fock (HF)\cite{Roothaan.Rev.Mod.Phys.1951,Pople_NesbetJ.Chem.Phys.1954} and second order M\o ller-Plesset perturbation theory (MP2),\cite{MP2.1934} the use of CD was expanded for a whole toolbox of methods which enabled a plethora of studies on molecules with thousands of basis functions: The spectrum ranges from ground-state scaled opposite-spin MP2 (SOS-MP2)\cite{Aquilante.Chem.Phys.Lett.2007} to excited state equation-of-motion coupled cluster (EOM-CCSD).\cite{Krylov.2013} 
The single-reference methods were complemented by Cholesky decomposed complete active space self-consistent field CASSCF\cite{Koch.MCSCF.J.Chem.Phys.2008} and second-order perturbation theory CASPT2\cite{Aquilante.2008} as well as the quadratically convergent implementations of SCF\cite{Lipparini.2021} and CASSCF.\cite{Lipparini.J.Chem.Theor.Comput.2021}
Besides developments that enable the efficient calculation of single-point energies, developments also include the calculation of properties via the implementation of CD for nuclear gradients\cite{Pedersen.J.Chem.Phys.2008,Pedersen.Int.J.Quantum.Chem.2013,Pedersen.J.Chem.Phys.2015,GaussJ.Chem.Phys.2019} at various levels of theory as well as CD for MP2 nuclear magnetic resonance shieldings.\cite{Burger.2021} Thus, using CD, studies for systems with more than thousand basis functions are these days readily available. 
\\
The situation is still somewhat different when turning to quantum-chemical predictions for molecules in finite magnetic fields. Quite a significant amount of finite-field (ff) quantum-chemical methodologies have been developed over the last years\cite{TellgrenJ.Chem.Phys.2008,Tellgren.Phys.Chem.Chem.Phys.2012,Tellgren.Science2012,Teale.J.Chem.TheoryComput.2015,Reynolds.Phys.Chem.Chem.Phys.2015,Stopkowicz.2015,Hampe.J.Chem.Phys.2017,Reimann.J.Chem.TheoryComput.2017,Teale.J.Chem.TheoryComput.2017,Stopkowicz.Mol.Phys.2018,ReynoldsJ.Chem.Phys.2018,GHF1,GHF2,Hampe.J.Chem.TheoryComput.2019,Tellgren.J.Chem.TheoryComput.2019,Lehtola.Mol.Phys.2019,Dichroism2019,HampePhys.Chem.Chem.Phys.2020,Bischoff.Phys.Rev.A.2020,pausch.klopperMolPhys2020,TealeJ.Chem.TheoryComput.2021,Helgaker.J.Chem.Phys.I.2021,Helgaker.J.Chem.Phys.II.2021} which are applicable to small to medium-sized molecules.
Finite magnetic-field calculations are associated with a high computational cost since in the general case the integrals and wave function parameters become complex-valued. As a result the  hard disk and/or memory requirements increase by a factor of at least two and the computational cost of the multiplication formally increases fourfold. 
Furthermore, the two-electron integrals have only fourfold instead of eightfold permutational symmetry and hence twice as many two-electron integrals have to be stored as compared to the field-free case.
Additionally, the number of integrals that vanish due to symmetry decreases, since the point-group symmetry of a molecule is usually reduced by the axial magnetic-field vector.\cite{Schmelcher.1990} 
A further important point is that for molecules in a magnetic field of arbitrary orientation, gauge-origin invariance has to be ensured. In general the phase factor of the exact wave function changes under the gauge-origin transformation so that the properties of the observables remain independent of the choice of the gauge origin of the magnetic field. However, approximated wave functions do not naturally exhibit this transformation behaviour. 
Hence, there the results may depend on the selected gauge origin. To ensure gauge-origin independence for calculations
with arbitrarily oriented finite magnetic field London orbitals\cite{London} are used, which are also referred to as gauge-including atomic orbitals (GIAOs).\cite{TellgrenJ.Chem.Phys.2008}\\
As a consequence of these challenges a reduction in the computational cost when dealing with larger systems (or large basis sets) becomes indispensable. 
Irons et al.\cite{Teale.J.Chem.TheoryComput.2017} have worked on ways to improve the performance of the integral evaluation by a careful choice of the underlying algorithm based on the angular momentum of the basis functions. Reynolds \& Shiozaki\cite{Reynolds.Phys.Chem.Chem.Phys.2015} and Klopper \& Pausch\cite{pausch.klopperMolPhys2020} have employed the RI approach in the context of finite-magnetic field developments. 
An advantage is here that the auxiliary basis set can be chosen real, leading to lower storage requirements and higher permutational symmetry of the 3-index integrals. With these developments, calculations with more than 1000 basis functions have become feasible. As mentioned earlier, RI methods need pre-defined auxiliary basis sets. These sets have to date only been optimized for field-free calculations. 
While the accuracy seems quite decent for energies when uncontracted auxiliary sets are employed (under 0.5 kJ/mol for field strengths lower than   1$\B \approx 235 000 $T ), a systematic error control and improvement of accuracy is not trivial.
\cite{pausch.klopperMolPhys2020} 
It is hence desirable to be able to rigorously control the error in the energies, in particular when the energy is studied as a function of the magnetic field. 
In order to address these issues and to offer an alternative route for finite-field computations for larger systems, in this work we present the Cholesky decomposition of the two-electron integrals over London orbitals. Our work reveals that compression rates are particularly beneficial in calculations with an external finite magnetic field and significantly depend on field strength and orientation. 
First, a short overview of the theoretical background of CD for ERIs is given in section \ref{sec:theory}. Our implementation is presented in section \ref{sec:implementation}. Validation of the implementation and an investigation of the errors in integrals and total energies are discussed in section \ref{sec:Validation}. In section \ref{sec:linear_independent_CB} the influence of the magnetic field on the structure of the ERI tensor and the respective consequences for the decomposition are discussed. In section  \ref{sec:breakevenpoint} the break-even-point of the algorithm using CD as compared to the canonical implementation is investigated. Finally, we conclude in section \ref{sec:pointgroup} and \ref{sec:representative_calcs} with a discussion of the treatment of symmetry and representative parallelized calculations on extended systems with more than 2000 basis functions.

\section{CD over London Orbitals}\label{sec:CDoverGIAOs}
\subsection{Theory}\label{sec:theory}
An element of the ERI matrix $\bm{V}$ over London orbitals 
\begin{equation}
    \omega_{\mu}=\text{e}^{-i\textbf{k}\textbf{r}}\chi_{\mu}
    \label{eq:LondonAO}
\end{equation}
composed of a standard Gaussian $\chi_\mu$ centered at $\textbf{K}_\mu$ and a complex phase factor (in which $\textbf{k}=\frac{1}{2}\textbf{B}\times (\textbf{K}_\mu-\textbf{G})$, $\textbf B$ is the magnetic field and $\textbf{G}$ is the gauge origin) can be approximated by the CD\cite{LinderbergJ.Quantum.Chem.1977,KochJ.Chem.Phys.2003,Aquilante.Chem.Phys.Lett.2007,Koch.MCSCF.J.Chem.Phys.2008,Aquilante.2008,Pedersen.J.Chem.Phys.2008,Krylov.2013,Pedersen.Int.J.Quantum.Chem.2013,Pedersen.J.Chem.Phys.2015,KochJ.Chem.Phys.2019,GaussJ.Chem.Phys.2019,Lipparini.2021,Lipparini.J.Chem.Theor.Comput.2021,Burger.2021} as
 \begin{equation}\label{eq:Cholesky_Zweielektronenmatrix_AO_Basis}
    V_{\mu\nu\sigma\rho}=\left(\mu\nu|\sigma\rho\right)\approx\left(\mu\nu|\sigma\rho\right)_{\text{CD}}=\sum_{J=1}^{N_{\text{CH}}}L_{\mu\nu}^{J}L_{\rho\sigma}^{J*}\;
    \end{equation}
where $N_{\text{CH}} $ is the number of Cholesky vectors (CVs) and corresponds to the numerical rank of the decomposition and  $\mu,\nu,\sigma,\rho$ correspond to the indices of the corresponding London orbitals. $L_{\mu\nu}^J$ is an element of the CV $\bm{\ell}^J$, such that $\bm{V} =\sum\limits_J\bm{\ell}^J\bm{\ell}^{J\dagger}$. In the field-free case and in the case of linear molecules in a parallel magnetic field, the maximum dimension corresponds to $\frac{N(N + 1)}{2}$ with $N$ as the number of basis functions. 
For all other molecules and orientations relative to the external magnetic field the maximal dimension increases to $N^2$ because of the reduced permutational symmetry of the ERIs
    \begin{equation*}
    (\mu\nu|\sigma\rho)=(\sigma \rho|\mu \nu) = (\nu \mu |\rho \sigma)^* = (\rho \sigma |\nu \mu)^* 
    .
    \end{equation*}
Note that this implies that $L_{\mu \nu}^J \ne L_{\nu \mu}^{J*}$.    
The elements of the CVs can be determined iteratively by
\begin{equation}
    L^{J}_{\sigma\rho}=\frac{1}{\sqrt{D^{J}_{\mu\nu}}}\left[\left(\sigma\rho|\mu\nu\right)-\sum_{K=1}^{J-1}L^{K}_{\sigma\rho}L^{K*}_{\nu\mu}\right]
    \label{eq:cholvec}
\end{equation}
and the corresponding diagonal elements by
\begin{equation}
    D^{J}_{\mu\nu}=\left(\mu\nu|\nu\mu\right)-\sum_{K=1}^{J-1}L^{K}_{\mu\nu}L^{K*}_{\mu\nu}\;.
\end{equation}
Note that the diagonal of the ERI in an external magnetic field is defined as $(\mu\nu|\nu\mu)$ to ensure positive semi-definiteness. In the field-free case because of the equivalence $(\mu\nu|\nu\mu)=(\mu\nu|\mu\nu)$ the latter form is typically used. 
The CD follows a (partial) pivoting procedure\cite{KochJ.Chem.Phys.2003} in which in each iteration a new CD vector with index $J$ is assigned to the largest of all updated diagonal elements of the ERI matrix with indices $\mu$ and $\nu$.
By truncating the decomposition at the iteration where all remaining updated diagonal elements are smaller than a chosen threshold $\tau=10^{-\delta}$, where $\delta$ is the Cholesky parameter, the CD removes (approximate) linear dependencies among the columns of the ERI matrix and leads to a compact representation.\cite{LinderbergJ.Quantum.Chem.1977}

Thus the number of required vectors is significantly smaller than the number of vectors for the full decomposition and the storage of the full integral matrix reduces from $\mathcal{O}(N^4)$ to $\mathcal{O}(N^2N_{\text{CH}}) $ for the Cholesky decomposition.\cite{LinderbergJ.Quantum.Chem.1977} 
The error for a given ERI vanishes if the numerical rank corresponds to the maximum dimension. In all other cases the error is determined by the choice of the threshold since all remaining diagonal elements are smaller than the tolerance criterion $\tau$. According to the Cauchy-Schwarz inequality 
\begin{equation}
    \left(\mu\nu|\sigma\rho\right)_\text{CD}\leq\sqrt{\left(\mu\nu|\nu\mu\right)_\text{CD}}\sqrt{\left(\sigma\rho|\rho\sigma\right)_\text{CD}}\leq\tau
\end{equation}
this upper bound holds for all remaining integrals as well and thus the error of the approximated ERI is strictly below the threshold. By choosing $\delta $ a reduction in the computational requirements together with a rigorous control of the error is hence possible.\\
For the iterative solution of the HF equations, the Fock matrix is built by substituting the two-electron integrals with the expression in Eq. \eqref{eq:Cholesky_Zweielektronenmatrix_AO_Basis} leading to a scaling of $\mathcal{O}(N_\mathrm{CH}N^2 O)$ with $O$ as the number of occupied orbitals.\cite{KochJ.Chem.Phys.2003}
For correlated calculations a sequential transformation of the CVs into the basis of molecular orbitals (MOs)
\begin{equation}\label{eq:transformierte_Cholesky_Vektoren}
    L_{pq}^{J}=\sum_{\mu\nu}C_{\mu p}^{*}L_{\mu \nu}^{J}C_{\nu q}
\end{equation}
is performed which replaces the four-index integral transformation by a two-index transformation of a limited set of vectors reducing the formal scaling to $\mathcal{O}(N^3N_{\text{CH}})$.\cite{LinderbergJ.Quantum.Chem.1977,Krylov.2013} The corresponding MO integrals are expressed analogously as
\begin{equation}\label{eq:Cholesky_Zweielektronenmatrix_MO_Basis}
    \left(pq|rs\right)\approx \sum_{J}^{N_{\text{CH}}}L_{pq}^{J}L_{sr}^{J*}
\end{equation}
and may be used in a subsequent post HF theories. In open-shell systems the MO integrals that do not vanish after spin integration can be represented by two sets of MO-CVs that occur exclusively for the spin pairs $\alpha\alpha$ and $\beta\beta$. 

\subsection{Implementation}\label{sec:implementation}

A complex CD routine which employs a partial pivoting algorithm\cite{KochJ.Chem.Phys.2003,KochJ.Chem.Phys.2019} has been implemented within the Mainz INTegral (MINT) package\cite{MINT} of the program package CFOUR\cite{cfour,cfour.2020} which uses the McMurchie–Davidson scheme\cite{McMurchieJ.Chem.Phys.1978,TellgrenJ.Chem.Phys.2008} for computing
integrals. The realization of ff-CD-HF follows the strategies described in reference \onlinecite{Lipparini.2021}. Additionally, ff-MP2 has been implemented. Both a closed-shell as well as an open-shell implementation based on unrestricted HF has been carried out. 
Summarizing the essential points of the implementation of CD with ff-HF and -MP2 methods we note:
\begin{itemize}
    \item The algorithms use complex data types.
    \item For the CD, the diagonal in Eq. \eqref{eq:cholvec} should be defined as $(\mu\nu|\nu\mu)$. 
    \item To increase computational efficiency a Cauchy-Schwarz-screening\cite{Ahlrichs.J.Comp.Chem.1989} has been implemented for the integral evaluation during the CD. 
    \item Point-group symmetry is exploited in the implementation. 
    \item Due to the fourfold permutational symmetry in the general case, the CVs are calculated and stored in symmetry-blocked fashion with $N^2$ values per vector. 
    More concretely, for the handling and storage of the CVs, it is used that the irreducible representation of a vector $\Gamma^{J}$ is determined by the symmetry of the direct product of the index pair $\overline{\mu\nu}$ of the maximum diagonal element. 
    This means that the vector of the corresponding integral column $\left(\sigma\rho|\overline{\mu\nu}\right)$ has a non zero entry if $\Gamma^J=\Gamma^{\bar{\mu}}\otimes\Gamma^{\bar{\nu}*}=\Gamma^{\rho}\otimes\Gamma^{\sigma*}$. 
    Hence, for totally-symmetric CVs ($\Gamma^J=\Gamma^\rho\ \otimes\Gamma^{\sigma*} = \Gamma^1$)  a square matrix for each irreducible representation is stored while for non-totally symmetric CVs, a  rectangular matrix with $N_{\Gamma^{\sigma}} \times N_{\Gamma^{\rho}}$ values is stored. 
    Note that in the field-free case, the storage of the blocks with $\Gamma^{\sigma}$ < $\Gamma^{\rho}$ is sufficient.
    \item Special cases, like $B=0$ or the case of linear molecules in a parallel magnetic field are handled separately as they allow to exploit the symmetries present in standard field-free calculations.
    \item An out-of-core algorithm was implemented which is used when not all CVs can be kept in memory. In such a case, the excess vectors that do not fit into the memory are written on disk and read when needed.
    \item For a faster convergence of the SCF wave function a DIIS\cite{PulayChem.Phys.Let.1980,PulayJ.Comp.Chem1982} scheme was employed.
    \item In the matrix multiplications for the build of the CVs, their transformation to the MO basis, and the assembly of the MO integrals, calls to efficient BLAS routines (ZGEMM) have been employed. With their use, multi threading via OpenMP\cite{OpenMP} can be employed in a simple manner by using respective threaded BLAS libraries.
    \item 
    In the calculation of the MP2 energy, the integrals of the type $(ov|ov)$ are needed where $o$ denotes indices of occupied and $v$ indices of virtual orbitals. $L_{ov}^J$ and $L_{vo}^J$ are calculated to reconstruct the integrals in the MP2 energy expression. 
    The reconstruction is performed in such a manner that there is only one symmetry block in memory at a time.
\end{itemize}

\begin{figure}[htbp]
    \centering
     \includegraphics[width=\linewidth]{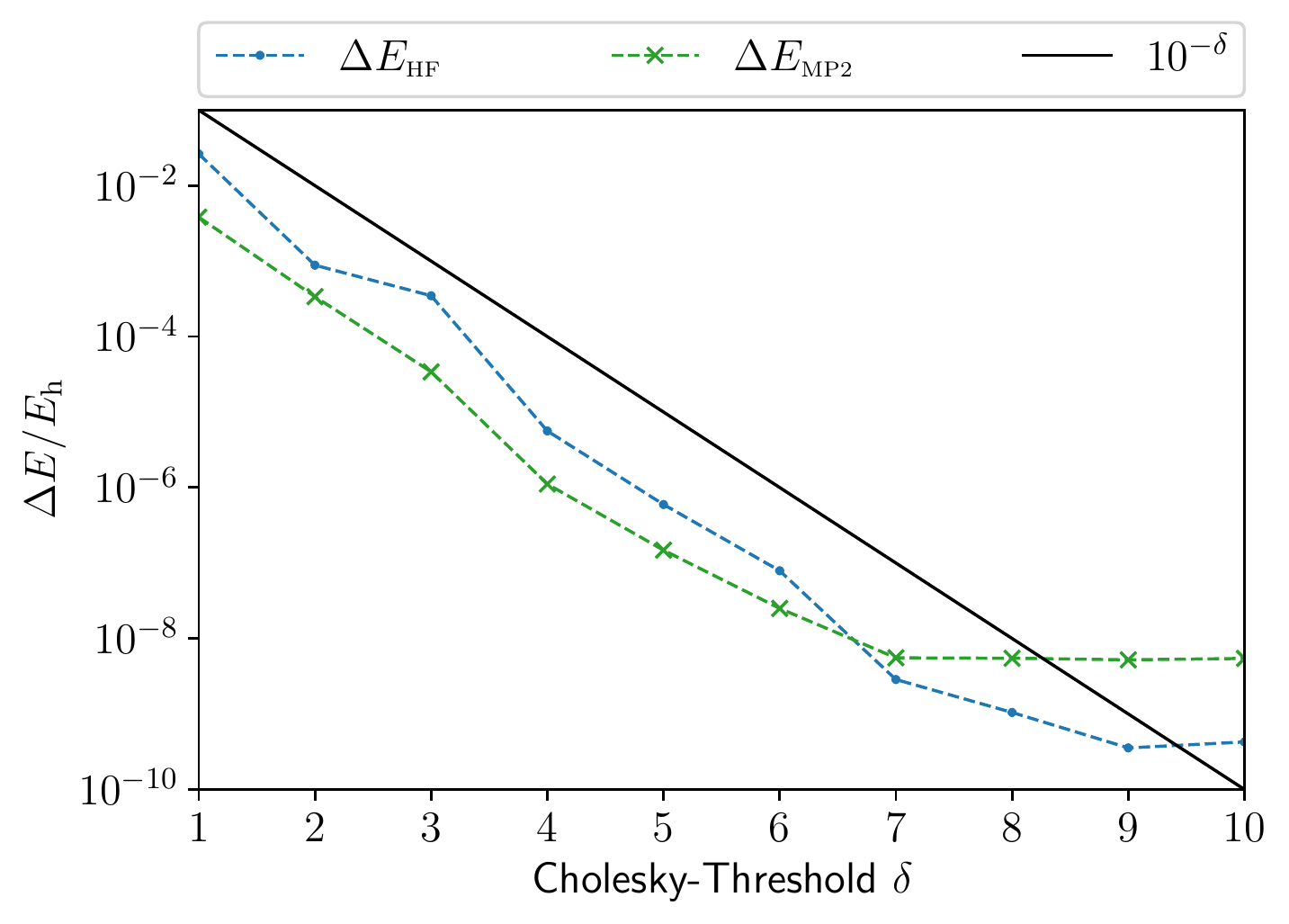}
    \caption{Mean error of the ff-CD-HF (blue) and ff-CD-MP2 (green) energy for the systems listed in table \ref{tab:system_geometries} as a function of the Cholesky parameter $\delta$. The SCF convergence was set to 10$^{-7}$.}
    \label{fig:meanerror}
\end{figure}

\section{Results and Discussions}
\subsection{Computational Details}

If not stated otherwise all calculations have been performed with the CFOUR\cite{cfour} program package using the uncontracted (unc) augmented (aug) versions of the correlation consistent Dunning\cite{Dunning.cc,Dunning.aug} or the Karlsruhe\cite{Karlruhe1992} basis sets. In calculations for the methylidyne radical (CH), water (H$_2$O), and ethylene (C$_2$H$_4$) Cartesian Gaussians were employed, the other calculations were performed with spherical Gaussians. The calculations have been carried out on an Intel(R) Xeon(R) E5-2643 node running at 3.40GHz. In section \ref{sec:representative_calcs}
a  Intel(R) Xeon(R) Gold 5215M node running at 2.50GHz was used. Parallel calculations used a total of 12 CPUs.
\subsection{Validation and Accuracy}\label{sec:Validation}

We validated the decomposition by reproducing the original ERIs from the CVs according to Eq. \eqref{eq:Cholesky_Zweielektronenmatrix_AO_Basis}. 
If the tolerance criterion is chosen to be very tight, the original ERI can be restored. This is possible for thresholds up to $10^{-12}$, as discussed in Ref. \onlinecite{LinearScalingTechniques}.   
The HF-SCF and MP2 codes were validated by comparing the energies with those of the existing program codes in CFOUR\cite{cfour} and LONDON.\cite{Londonprogramm,TellgrenJ.Chem.Phys.2008} 
Additionally, we verified the gauge-origin independence of the results.\\
To investigate the remaining error in the energy for truncated CDs, the closed-shell systems water (H$_2$O), ethylene (C$_2$H$_4$), and the open-shell methylidyne radical (CH) were calculated for various Cholesky thresholds. 
The systems were placed in a magnetic field with a strength of $1\B$ ($1\B \approx 235 000 $T) with three relative orientations that for simplicity we label as parallel ($||$), perpendicular ($\perp$) and 'arbitrary' ($/$) with respect to the magnetic-field vector. 
An overview of the calculated structures with respective orientations with respect to the magnetic field is given in table \ref{tab:system_geometries}. 
The mean error at the ff-CD-HF and ff-CD-MP2 levels for different Cholesky parameters compared to reference calculations without Cholesky decomposition is depicted in figure \ref{fig:meanerror}.
For small Cholesky thresholds the mean error in the HF (blue) and the MP2 (green) energies decreases nearly linearly with the Cholesky parameter up to a value of $\delta = 7$. The magnitude of the mean errors is strictly below the chosen Cholesky threshold - depicted by the black line ($ \Delta E_{\text{HF}} <10^{-\delta} $). 
These findings are in line with the observations for field-free HF calculations, i.e., the accuracy of the integrals, fixed rigorously via the Cholesky threshold, translates also to the SCF energy.\cite{KochJ.Chem.Phys.2003} 
For large thresholds the ff-HF error plateaus and eventually becomes larger than the threshold of the decomposition. 
This finding is explained simply by the fact that the SCF convergence of the orbital coefficients was chosen to be only $10^{-7}$. This choice limits the accuracy of the orbitals and hence also of the total energy. 
We tested this conclusion by increasing the SCF convergence for these cases which indeed leads to errors below the black line. 
For MP2, a similar trend is encountered (for the same reasons) except that the error introduced by setting the SCF convergence is more pronounced due to the linear dependence of the energy on the error in the orbital energies in the MP2 energy denominator.\cite{KochJ.Chem.Phys.2003}\\
For a comparison reference \onlinecite{pausch.klopperMolPhys2020} reports errors of the magnitude of $10^{-4}$ Hartree for finite-field generalized HF\cite{GHF1,GHF2} energies when employing an RI scheme using an uncontracted def2 auxiliary basis set which was optimised for calculations without magnetic fields. 
This accuracy can already be reproduced by using a Cholesky parameter of $\delta=4$ which is consistent with the benchmark study for field-free calculations done in reference \onlinecite{Aquilante.Chem.Phys.Lett.2007}. Important advantages of the CD are that the accuracy of the CD can be chosen a priori and that the CD is system-specific such that the errors of the CD are independent of the magnetic-field strength (see SI for details).

\begin{table}[htbp]
\centering
\caption{Calculated systems, employed basis sets, and orientations of the molecules and the magnetic-field vector (of strength $B$=1.0$\B$) in the chosen coordinate system. The field-free geometries for ethylene and the methylidyne radical were employed which were calculated at the CCSD/unc-pVTZ and unc-aug-pVDZ levels of theory, respectively. For water, the experimental geometry (ROH=0.958 \AA, HOH=104.45$^\circ$) from Ref. \onlinecite{Wassergeo} was used. }
\label{tab:system_geometries}
\begin{ruledtabular}
\begin{tabular}{@{}ccr@{}}
Molecule & Basis & Orientation  \\ \midrule
\multirow{3}{*}{Water (xy)}       & \multirow{3}{*}{unc-cc-pVQZ}                & y ( || )              \\
         &       & z ($\perp$)       \\
         &       &  45$^\circ$yz ( / )      \\
         \midrule
\multirow{4}{*}{Ethylene (xy)}      & \multirow{4}{*}{unc-aug-cc-pVTZ}            & x ( || )         \\
         &       & y ($\perp$)       \\
         &       & z ($\perp$)       \\
         &       & 45$^\circ$xz ( / )       \\
         \midrule
\multirow{3}{*}{Methylidyne (z)} & \multirow{3}{*}{\parbox{2.5cm}{unc-aug-cc-pVXZ\\ X=D,T,Q,5}} & z ( || )  \\
         &       & x ($\perp$)          \\
         &       & 45$^\circ$xz ( / )        \\ 
\end{tabular}%
\end{ruledtabular}
\end{table}

\subsection{Dependence on magnetic-field strength and orientation}\label{sec:linear_independent_CB}
It is well-known that the compression rate, i.e., the quotient between the full and the numerical rank of the decomposition, depends on the following factors: 
\begin{enumerate}
    \item It increases for large or uncontracted and diffuse basis sets, since the number of linear dependencies within the basis set grows with the number of basis functions.\cite{Koch.J.Chem.Phys.2008,Pedersen.Theor.Chem.Acc.2009} This is of relevance here because uncontracted and diffuse basis sets are typically needed in ff quantum-chemical calculations in order to adequately describe the anisotropy of the orbitals. 
    \item It increases with the system size. For large systems the compression rates grow because of the increased sparsity in the ERIs where the latter is a result of the decay with the inverse of the separation of the charge distributions.\cite{DyczmonsTheoret.Chim.Acta1973,Ochsenfeld.J.Chem.Phys.2005,Ochsenfeld.J.Chem.Phys.2005erratum} 
    \item It depends in a non-trivial manner on the molecular geometry.\cite{Pedersen.Theor.Chem.Acc.2009}
\end{enumerate}
In this section, we investigate how the compression rate changes when varying the strength of the magnetic field and its orientation  with respect to the molecule. 
Here, we study the methylidyne radical (CH) calculated at the ff-CD-MP2/unc-aug-cc-pVQZ with a fixed bond length (see table \ref{tab:detailedtimings} for details).  

\begin{figure*}[htbp]
     \centering
     \begin{subfigure}[b]{0.44\linewidth}
         \centering
         \includegraphics[width=\linewidth,trim={10.5cm 5.5cm 11.5cm 4.5cm},clip]{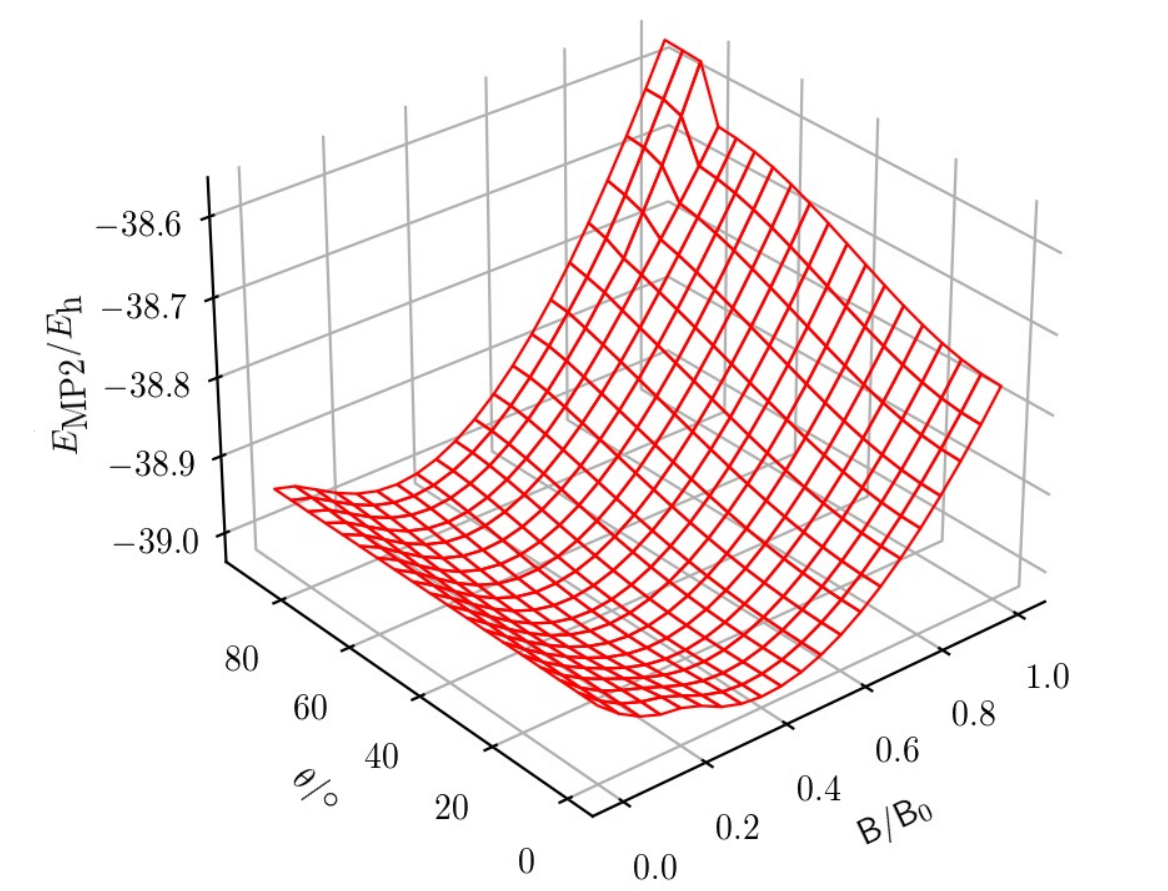}
         \caption{Total MP2 energy}
         \label{fig:E}
     \end{subfigure}
     \hfill
     \begin{subfigure}[b]{0.44\linewidth}
        \centering
         \includegraphics[width=\linewidth,trim={11cm 4.5cm 11cm 5.5cm},clip]{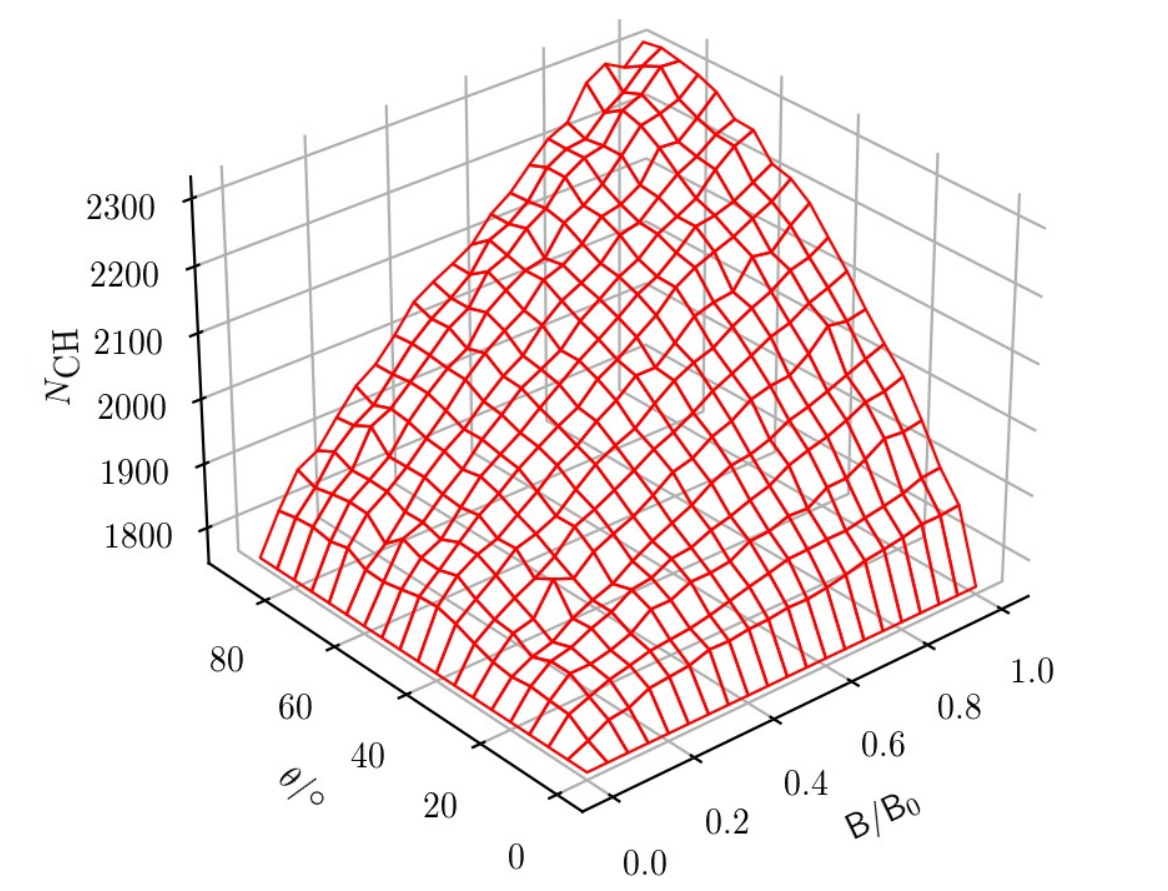}
         \caption{Number of CD vectors}
         \label{fig:Nch}
     \end{subfigure}
     \hfill
    \caption{Potential hypersurface \ref{fig:E} 
     of the methylidyne radical (CH) calculated at the ff-CD-MP2/unc-aug-cc-pVQZ level of theory and the respective numerical rank \ref{fig:Nch} of the CD with a constant Cholesky parameter of $\delta=9$ as a function of the magnetic-field strength $B$ and the angle $\theta$ between C-H bond axis and magnetic-field vector.}
    \label{fig:potentialhypersurface}
\end{figure*}
Figure \ref{fig:potentialhypersurface} shows the dependence of the MP2 energy and the number of CVs of the methylidyne radical (CH) on the magnetic-field strength $B$ and the angle $\theta$ between the magnetic-field vector and C-H bond axis. 
The edges of the potential surface define a magnetic field of strength $B=0\B$ or a parallel magnetic field of strength $B=0-1.0\B$. 
At these edges, the ERIs exhibit eightfold permutational symmetry and the number of CVs is constant due to the definition of the London orbitals in Eq. \eqref{eq:LondonAO}.

Trivially, the exponent of the phase factor \textbf{k} and thus the dependence on the magnetic field vanishes for molecules in the field-free case $ B = 0 \B$. This also applies to linear molecules in parallel magnetic fields, i.e., with the bond and the field oriented along the $z$-axis, because 
\begin{equation}\label{eq:Kreuzprodukt_phasenfaktor}
    \textbf{k}=\frac{1}{2}\textbf{B}\times (\textbf{K}_{\mu}-\textbf{G})=\frac{1}{2}
    \begin{pmatrix}
    0 \\
    0\\
    B_z
    \end{pmatrix}
    \times
    \begin{pmatrix}
    0 \\
    0\\
    K_{\mu,z}-G_z
    \end{pmatrix} = \mathbf 0.
\end{equation}
In these cases the London orbital corresponds to the field-free Gaussian orbital and the ERIs are real and independent of the magnetic field.\footnote{The total energy is obviously still dependent on the magnetic-field strength via to the paramagnetic and diamagnetic terms in the Hamiltononian as seen in figure \ref{fig:E}. 
For a detailed discussion on the symmetry of the ERI over London orbitals in a finite magnetic field see Ref. \onlinecite{ComplexEn21}.} 
As a result the ERIs do not change with the magnetic field and the numerical rank of the decomposition remains constant. 
For all other orientations of the molecule in a magnetic field, the permutational symmetry is reduced from eight to four. Figure \ref{fig:Nch} shows the increase in the number of CVs with the magnetic-field strength $B$ and the angle $\theta $, with a maximum at an angle of $\SI {90}{\degree}$ and a magnetic-field strength of 1$\B$. 
Considering the following properties of the CD that:\cite{LinderbergJ.Quantum.Chem.1977,Pedersen.J.Chem.Phys.2008,Pedersen.Theor.Chem.Acc.2009,  LinearScalingTechniques}
\begin{enumerate}
    \item the CD is equivalent to a modified Gram-Schmidt orthonormalization of the product densities $ |\mu\nu) $,
    \item the CD eliminates the linear dependencies in the basis set up to a tolerance criterion of $10^{-\delta}$. The functions $ |\mu\nu) $ chosen by the decomposition in each iteration span a linearly independent, orthonormal Cholesky basis, 
    \item the Cholesky basis and the given AO basis span the same space, 
    \item the Cholesky basis is the best possible auxiliary basis in an RI context, 
\end{enumerate}
the increase in the numerical rank of the decomposition with the magnetic-field strength and angle can be understood as a decrease in linear dependencies in the space of the atomic orbital densities. For example, due to the permutational symmetry in the ERIs two columns are linearly dependent in the sense of the Cholesky decomposition if
\begin{equation}
    (\mu\nu|\sigma\rho)-(\mu\nu|\rho\sigma) < 10^{-\delta}\;.
    \label{eq:numdiff}
\end{equation}
Since this difference vanishes for integrals over GTOs it is entirely determined by the phase factors of the London orbitals. 
Another look at the wave vector $\textbf{k} $ in Eq. \eqref{eq:Kreuzprodukt_phasenfaktor} shows that first of all the difference scales with the magnetic-field strength $B$. 
Second, the cross product of two vectors ($ \textbf{a} \times \textbf{b} = \abs {\textbf{a}} \abs{\textbf{b}} \sin{\theta} $) maximises for an angle of \SI{90}{\degree} and equally  the difference in Eq. \eqref{eq:numdiff} scales with the angle between magnetic field and bond axis. 
As a result, product densities in the ERI that are linearly dependent in the field-free case become more and more clearly linearly independent with increasing angle and magnetic-field strength. 
Since the CD removes linearly dependent product densities below the Cholesky threshold and keeps orthonormal ones this trend causes the Cholesky basis to grow, i.e., the number of CVs to increase. 
In our example, the number of basis functions is 175 and hence the theoretical maximum dimension would be 15400 and 30625 for the field-free and field-dependent CVs, respectively.
In the calculations, for $\delta=9$, in the field-free or parallel orientation, the number of CVs is 1756. 
The largest numbers of CVs is 2315, encountered at a perpendicular magnetic-field orientation and the largest considered field strength of 1 $\B$. 
The largest compression rates are hence found for small field strengths and near-parallel orientations. 
Here, the compression rate is 17.44 for a close-to-parallel and near-field-free configuration as compared to 13.22 for the perpendicular orientation at 1 $\B$. For the field-free case the compression rate here is 8.77.
This is because when transitioning from a field-free case or a linear system in a parallel magnetic field to a different orientation, the formal maximum dimension is nearly doubled. For the slightly tilted orientations with respect to the magnetic field, it holds that 
elements of the ERI tensor that are exactly equal by symmetry in the field-free or parallel orientation are no longer equal -~but still very similar numerically in the sense of the difference in Eq. \eqref{eq:numdiff}.
The CD now removes all contributions that are linearly dependent numerically, leading to large compression rates and a small number of CVs.
Note that in our example, the Cholesky threshold is much larger ($\delta=9$) than what would be used in normal calculations and hence much larger compression rates would be observed there. 

Due to the innate connection between RI and CD\cite{LinderbergJ.Quantum.Chem.1977,KochJ.Chem.Phys.2003} information about the Cholesky basis may support the generation of magnetic field-dependent auxiliary basis sets for an RI treatment. 
We note that the number of required Cholesky vectors changes quite significantly not only with the strength but also with the orientation of the magnetic field, which should then probably be taken into account. 
As the Cholesky algorithm automatically and in a black-box manner generates the required number of CVs to retain a certain accuracy, without having to rely on a fixed pre-defined auxiliary basis set, CD is particularly useful in the context of finite magnetic-field calculations with varying field strengths and orientations.

\subsection{Break-even point}\label{sec:breakevenpoint}

\begin{figure}
    \centering
    \includegraphics[width=\linewidth,trim={11.5cm 5cm 11.5cm 2cm},clip]{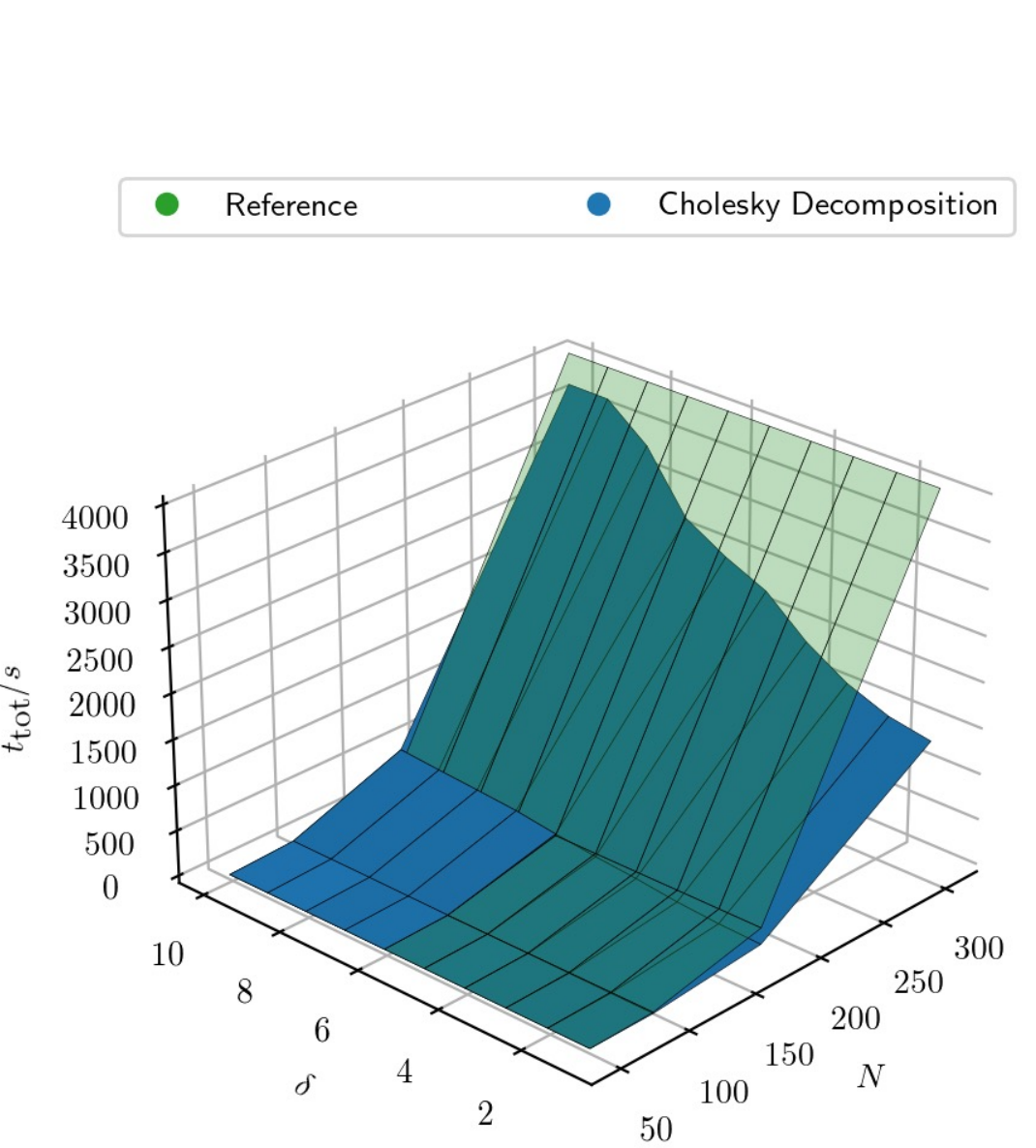}
    \caption{Total wall clock timings for calculations on the methylidyne radical (CH) at the ff-MP2/unc-aug-cc-pVXZ (X=D,T,Q,5) level. Green: Conventional ff-MP2 reference calculations, green: ff-CD-MP2 calculations performed at a magnetic-field strength of $1.0\B$ and an angle of 45$^\circ$ between the C-H bond axis and the magnetic field and $\delta=1-10$. 
    }
    \label{fig:breakevenpoint}
\end{figure}

An important motivation in using CD is the reduction of memory requirements and the speed-up of quantum-chemical calculations.
When using CD, the theoretical scaling is a function of the number of CVs, the latter being significantly smaller in actual calculations than the maximum dimension of the full decomposition. 
For a given molecule, the number of CVs generally depends on the chosen Cholesky threshold and the basis-set size. To assess the break-even point between the implementations using the full and the decomposed ERI, the timings of calculations on the methylidyne radical are plotted in figure \ref{fig:breakevenpoint} as a function of the Cholesky parameter $\delta$ and the size of the basis set. 
All calculations were performed with CFOUR for a field strength of 1 $\B$ and an angle of 45$^\circ$ between the bond axis and the magnetic field. 
The application of the CD results first of all in an acceleration for parameters $\delta\leq5$ independent of the basis-set size. Those thresholds are already able to reproduce the energies to an accuracy of up to $\approx10^{-7}$ Hartree. This error margin is even smaller than the typical basis-set error.\footnote{Extrapolation of the HF energy to the basis set limit yields an error of $1.7\cdot10^{-4}E_{\text{h}}$ with respect to the unc-aug-cc-pV5Z basis set.} 
Secondly, the CD computations are faster than their respective standard ff counterparts for basis sets with more than 175 basis functions since here generally higher compression rates are observed. 

In table \ref{tab:detailedtimings} a detailed comparison of the timings is listed for computations on the methylidyne radical CH, benzene, boric acid and the cyclopentadienyl anion CP$^-$. 
While the integral evaluation is the bottleneck of the conventional calculations for small systems, the MO transformation starts to govern the computing time for larger systems due to the formal $N^5$ scaling with the basis-set size. 
For the CD the timings for the integral evaluation may increase due to the recalculation of integral batches which is a consequence of the partial-pivoting algorithm which makes the integral evaluation the time-determining factor. 
Nonetheless, the decomposition leads to a speed-up of the SCF procedure and the MO transformation which scale with $\mathcal{O}(N_{\text{CH}}N^2O)$  and $\mathcal{O}(N_{\text{CH}}N^3)$, respectively.\cite{KochJ.Chem.Phys.2003} 
Using the Cholesky decomposed ERIs compared to the full ERI tensor is slower only for very small basis sets and tight thresholds which would not be used in actual calculations. 
For all other cases the Cholesky scheme provides a significant speed-up while at the same time maintaining a reasonable and controllable accuracy.

\begin{table*}[htbp]
\centering
\caption{Detailed comparison of timings\protect\footnote{The discrepancy between the sum of all timings and the total timings are due to not listed setup and I/O timings.} between calculations with and without the use of CD ($\delta=5$). The table shows timings for a) the methylidyne radical at the ff-MP2/unc-aug-cc-pVXZ (X=D,T,Q,5) levels with the magnetic-field vector of $1.0\B$ being tilted at a 45$^\circ$ angle to the C-H bond axis and b) the molecules benzene, boric acid, cyclopentadienyl anion (CP$^-$), and staggered ethane at the  ff-MP2/cc-pVTZ level. For the first three molecules the magnetic-field vector is oriented perpendicular to the molecular plane. For ethane, the magnetic-field vector is oriented along the C-C bond axis. The geometries were obtained from respective field-free optimizations at the CCSD/cc-pVTZ level.}
\label{tab:detailedtimings}
\begin{ruledtabular}
\begin{tabular}{@{}lccccccclccccccc@{}}
 a) & X & $t_{\text{int}}$/s  & $t_{\text{chol}}$/s\protect\footnote{\label{note:tchol}For the timings of $t_{\text{chol}}$ only the build of the CVs as in equation \eqref{eq:cholvec} is considered which is the most expensive step of the Choleksy procedure scaling $\mathcal{O}(N^2N_{\text{CH}}^2)$.}
 & $t_{\text{HF}}$/s\protect\footnote{\label{note:HFpit}Timings 'per iteration' are given in parentheses.}  &$t_{\text{ao2mo}}$/s & $t_{\text{MP2}}$/s   & $t_{\text{tot}}$/s & b) & Molecule & $t_{\text{int}}$/s  & $t_{\text{chol}}$/s$^{\text{\ref{note:tchol}}}$ & $t_{\text{HF}}$/s$^{\text{\ref{note:HFpit}}}$  &$t_{\text{ao2mo}}$/s & $t_{\text{MP2}}$/s   & $t_{\text{tot}}$/s           \\\midrule
\multirow{4}{*}{Ref.} & D & 0.34  & -      & 0.49 (0.03)   & 0.26    & 0.001 & 1.24  & \multirow{4}{*}{Ref.} & Benzene     & 1929 & -    & 434 (20)   & 236  & 0.49 & 2735 \\
                      & T & 6.63  & -      & 7.48 (0.42)   & 5.20    & 0.003 & 21    & & Boric acid   & 219  & -    & 48  (2.83) & 23   & 0.10 & 310  \\
                      & Q & 100   & -      & 90   (5.02)   & 93      & 0.01  & 300 &  & CP$^-$ & 916  & -    & 173 (9.60) & 96   & 0.24 & 1251 \\
                      & 5 & 1576  & -      & 864  (48)    & 1466    & 0.04  & 4069 & & Ethane      & 89   & -    & 23  (1.80) & 13   & 0.02 & 138  \\\midrule
\multirow{4}{*}{CD}   & D & 0.95  & 0.09   & 0.21 (0.01)   & 0.02    & 0.01  & 1.48  &\multirow{4}{*}{CD}   & Benzene     & 848  & 98   & 40 (2.49) & 8.20 & 9.27 & 1010 \\
                      & T & 16.09 & 1.29   & 1.81 (0.11)   & 0.23    & 0.06  & 20  &  & Boric acid   & 218  & 16   & 7.74 (0.52) & 1.29 & 1.28 & 247  \\
                      & Q & 248   & 228    & 13 (0.75)  & 2.26    & 0.35  & 265   & &CP$^-$ & 499  & 49   & 22 (1.29) & 3.90 & 3.91 & 613  \\
                      & 5 & 2005  & 177    & 72 (4)  & 19      & 1.84  & 2285  & &Ethane  & 68   & 7.69 & 3.74 (0.31) & 0.76 & 0.28 & 81 

\end{tabular}%
\end{ruledtabular}
\end{table*}

\subsection{Point-Group Symmetry}\label{sec:pointgroup}

\begin{figure}[htbp]
    \centering
    \includegraphics[width=\linewidth]{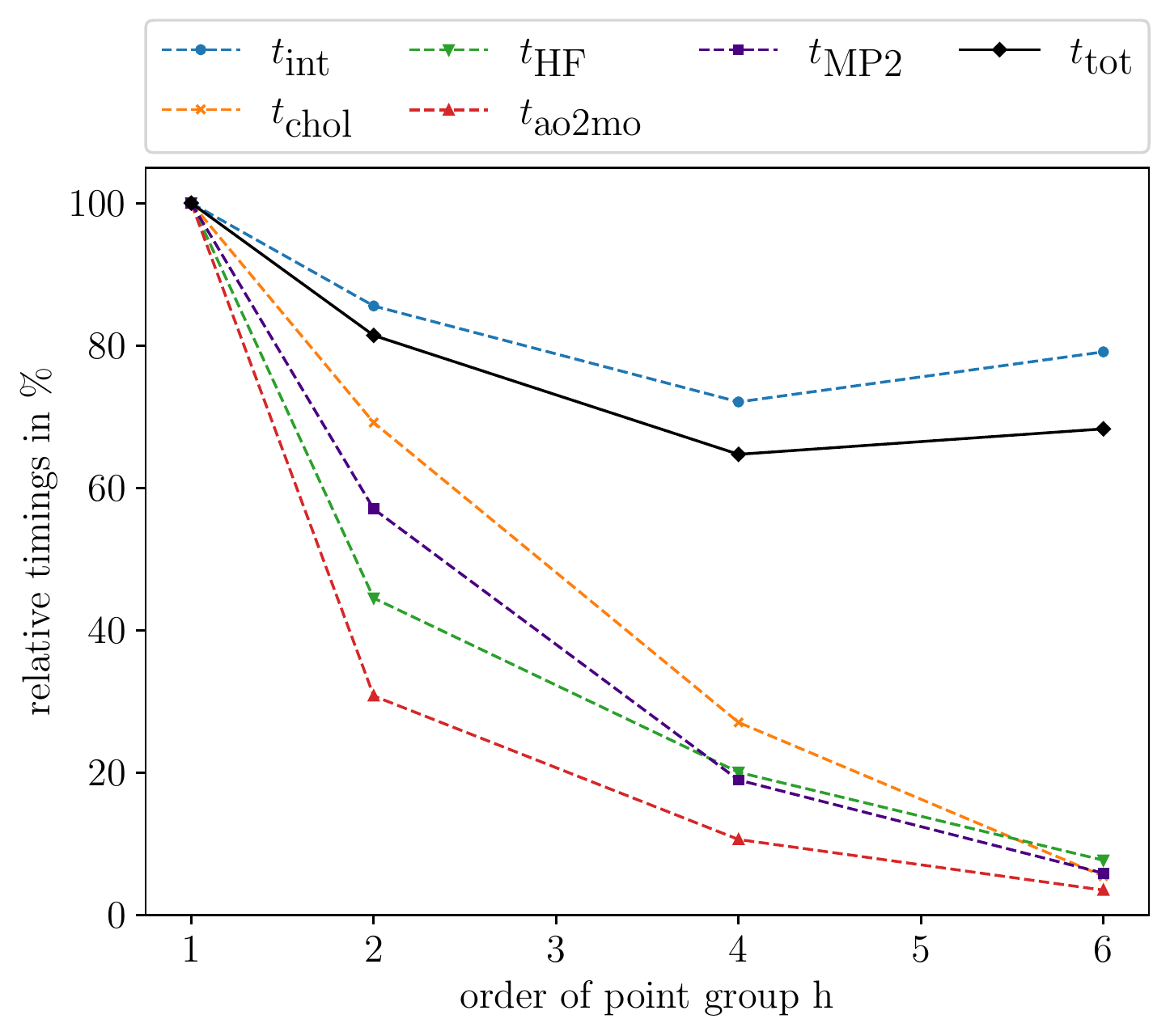}
    \caption{Relative (mean) timings for the integral evaluation (int), the build of the CVs (chol), the average time per Hartree-Fock iteration (HF), the MO transformation (ao2mo) and the computation of the MP2 energy as well as the total time (tot) for a calculation on benzene in a magnetic field of the strength 0.1$\B$ perpendicular to the molecular plane with the cc-pVTZ basis and $\delta$=5 as a function of the order of the computational point group. The corresponding point groups are the real $C_1$, $C_2/C_s$, $C_{2h}$ and the complex $C_{3h}$.}
    \label{fig:symsys}
\end{figure}

Additional computational speed-up can be achieved by the treatment of the point-group symmetry. 
In an external magnetic field, all axes of rotation perpendicular to the magnetic-field vector and all mirror planes that contain the magnetic-field vector no longer constitute valid symmetry elements. 
As a result, the axis of rotation parallel to the magnetic-field vector becomes the main axis and only mirror planes perpendicular to the magnetic-field vector are retained. Thus the point-group symmetry is typically reduced.\cite{Schmelcher.1990} 
The achieved relative speed-up through the exploitation of point-group symmetry in calculations using CD is shown in figure \ref{fig:symsys} for benzene in a finite magnetic field. 
The corresponding absolute timings are listed in table \ref{tab:absolute_timings_symsys}. 
In the field-free case the point-group symmetry is $D_{6h}$. In a magnetic-field vector oriented perpendicular to the molecular plane the symmetry is reduced and the largest real Abelian subgroup of the full molecular point group is $C_{2h}$. 
It is always possible to treat the system in a subgroup of $C_{2h}$, namely $C_{2}$ and $C_s$, at the cost of the order of the point group which decreases from 4 to 2. 
The timings of the calculation clearly show a faster performance relative to the reference calculations in $C_1$ that scales with the order of the point group. 
Thereby the wall time for $C_{2h}$ is roughly 70\% faster for the build of the CVs, the average time per HF iteration, the transformation of the integrals in the MO basis and the calculation of the MP2 energy. 
For the integral evaluation a smaller speed-up of only  28\% is achieved. 
This is also reflected in the total timings which are dominated by the integral-evaluation step as discussed in section \ref{sec:breakevenpoint} resulting in a net reduction of the total wall-clock time from 1010 seconds for $C_1$ to 653 seconds for $C_{2h}$. 
A further acceleration can be achieved by increasing the order of the point group which is possible by employing a complex Abelian point group.\cite{Petros.unpub} In this example the point group $C_{3h}$ ($h=6$) results in a speed-up of all steps except the integral evaluation\footnote{due to additional loops in the handling of the double-coset decomposition\cite{ESQC.Book1} in the integral evaluation, a slight increase in computation time is observed when a complex Abelian point groups is used. 
While not of very much use here in terms of computational speed, for other correlated methods such as those of the coupled-cluster type, a net reduction in computation time is expected since the integral evaluation is not the rate determining step anymore.} of about 95\% relative to a calculation in $C_1$. For example the build of the CD decreases from 98 seconds down to 5 seconds.

\begin{table}[htbp]
\centering
\caption{Detailed list of timings\protect\footnote{The discrepancy between the sum of all timings and the total timings are due to not listed setup and I/O timings.} (integral evaluation (int), the build of the CVs (chol), the average time per Hartree-Fock iteration (HF), the MO transformation (ao2mo) and the computation of the MP2 energy as well as the total time (tot)) for calculations on benzene at the ff-CD-MP2/cc-pVTZ level using spherical Gaussians with $\delta=5$ as a function of the order of the computational point group. The computational point groups are the real $C_1$, $C_2/C_s$, $C_{2h}$ and the complex $C_{3h}$.  }
\label{tab:absolute_timings_symsys}
\begin{ruledtabular}
\begin{tabular}{@{}lccccccc@{}}
 $h$ & Point Group & $t_{\text{int}}$/s  & $t_{\text{chol}}$/s\protect\footnote{For the timings of $t_{\text{chol}}$ only the build of the CVs as in equation \eqref{eq:cholvec} is considered which is the most expensive step of the Choleksy procedure scaling $\mathcal{O}(N^2N_{\text{CH}}^2)$.} & $t_{\text{HF}}$/s\protect\footnote{Timings 'per iteration' are given in parentheses.}  &$t_{\text{ao2mo}}$/s & $t_{\text{MP2}}$/s   & $t_{\text{tot}}$/s           \\\midrule
1 & $C_1$  & 848 & 98   & 40 (2.49) & 8.20 & 9.27 & 1010 \\
2 & $C_2$  & 632 & 50   & 14 (1.06) & 2.46 & 5.12 & 708  \\
2 & $C_s$  & 819 & 87   & 19 (1.16) & 2.58 & 5.45 & 936  \\
4 & $C_{2h}$ & 611 & 27   & 10 (0.50) & 0.87 & 1.76 & 653  \\
6 & $C_{3h}$ & 670 & 5.35 & 11 (0.19) & 0.28 & 0.54 & 689 
\end{tabular}%
\end{ruledtabular}
\end{table}

\subsection{Representative Calculations}\label{sec:representative_calcs}

\begin{figure*}
     \centering
     \begin{subfigure}[b]{0.3\linewidth}
         \centering
         \includegraphics[width=\linewidth,trim={4cm 2cm 4cm 2cm},clip]{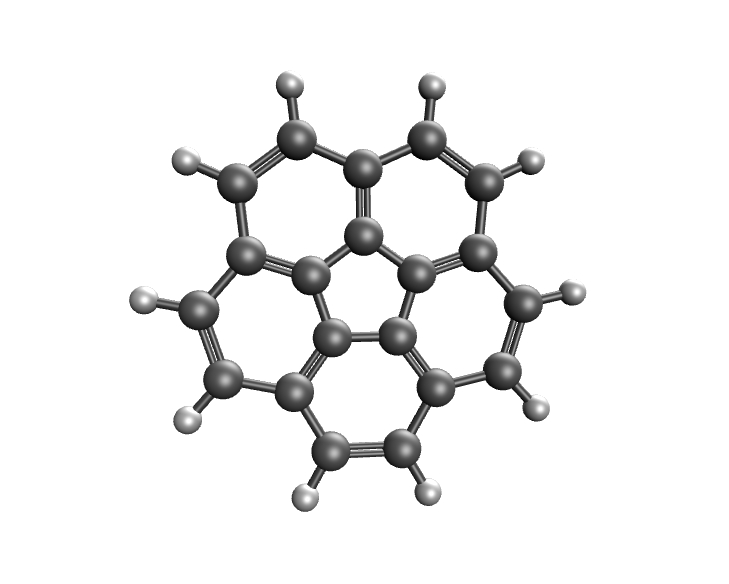}
         \caption{Corannulene}
         \label{fig:Corannulen}
     \end{subfigure}
     \hfill
     \begin{subfigure}[b]{0.3\linewidth}
         \centering
         \includegraphics[width=\linewidth,trim={4cm 1.5cm 4cm 1.5cm},clip]{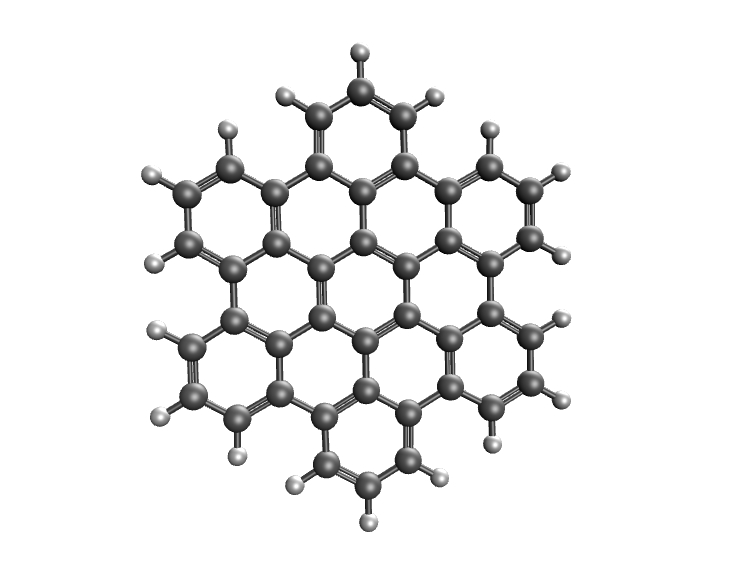}
         \caption{Hexabenzocoronene}
         \label{fig:HBC}
     \end{subfigure}
     \hfill
     \begin{subfigure}[b]{0.3\linewidth}
         \centering
         \includegraphics[width=\linewidth,trim={4cm 2cm 4cm 2cm},clip]{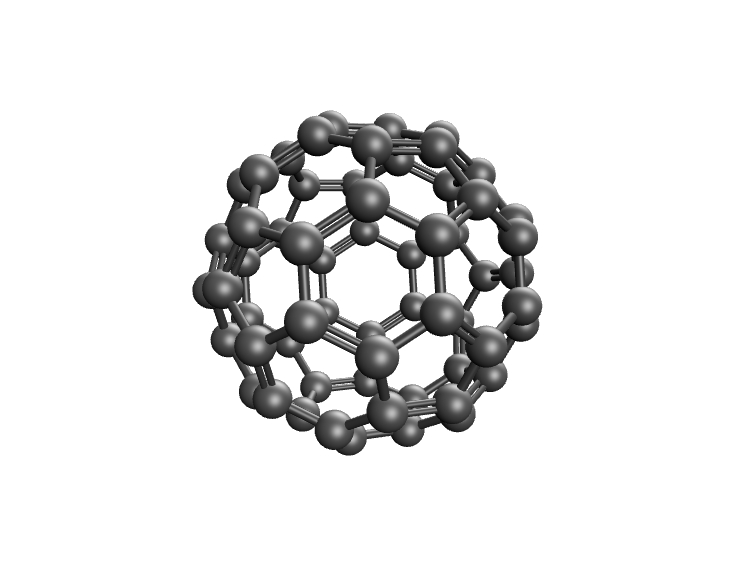}
         \caption{Fullerene}
         \label{fig:Fulleren}
     \end{subfigure}
     \hfill
     \begin{subfigure}[b]{0.3\linewidth}
         \centering
         \includegraphics[width=\linewidth,trim={4cm 2cm 4cm 2cm},clip]{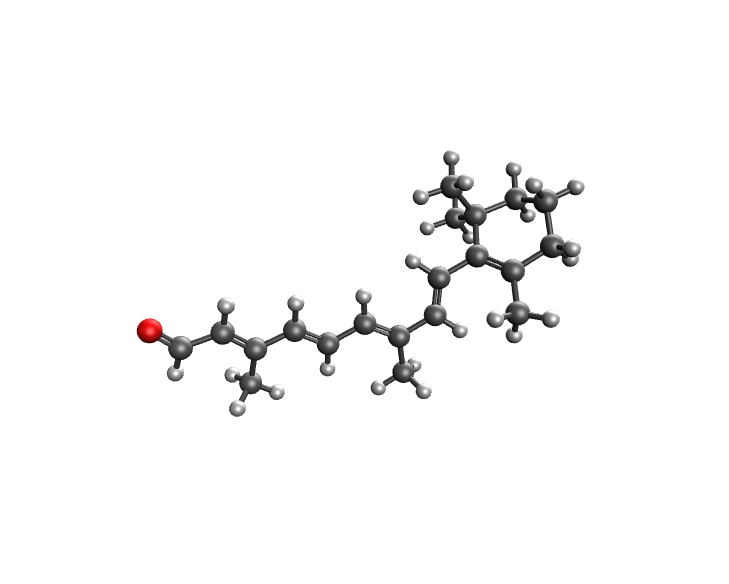}
         \caption{Retinal}
         \label{fig:Retinal}
     \end{subfigure}
    \caption{Molecular structure of calculated molecules. The colours scheme of the atoms correspond to: black = carbon, white = hydrogen, and red = oxygen.}
    \label{fig:Molecules}
\end{figure*}

In reference \onlinecite{TellgrenPhys.Chem.Chem.Phys2009} the paramagnetic-to-diamagnetic transition from closed-shell molecules with an extended $\pi$-system in a strong magnetic field has been studied. The largest calculation reported there was a ff-HF calculation performed on the corannulene dianion using a cc-pVDZ basis with 590 primitive basis functions. 
The authors pointed out that for larger systems, a paramagnetic-to-diamagnetic transition might occur for a relatively small critical field strengths that could be reproduced experimentally. 
Notably, in a magnetic field such a closed-shell paramagnetic state would quickly become an excited state while an open-shell state of higher multiplicity would likely become the ground state of the system. 
At the time, calculations on larger systems  were out of reach. But recent developments in the field\cite{Teale.J.Chem.TheoryComput.2017,pausch.klopperMolPhys2020} as well as the implementation reported here have made it possible to treat larger systems in a finite magnetic field.
While attempting a full study of the electronic structure and the paramagnetic-to-diamagnetic transition of such systems in a magnetic field is certainly an interesting topic for future studies, here we demonstrate the applicability of the present development to systems with up to 2000 basis functions. 
In figure \ref{fig:Molecules} the calculated molecules are listed starting from the corannulene dianion \ce{C20H10^2-}, hexabenzocoronene \ce{C42H18} (HBC), the buckminsterfullerene \ce{C60}, and retinal \ce{C20H28O}.
The field-free structures for HBC and \ce{C60} were taken from Refs. \onlinecite{OchsenfeldJ.Am.Chem.Soc.2001} and \onlinecite{HaeserChem.Phys.Lett.1991}, respectively. For corannulene and the retinal molecule, a field-free density-functional theory/B3LYP\cite{B3LYPa,B3LYPb} geometry was calculated with Turbomole\cite{TURBOMOLE,TURBOMOLE2} using a 6-31G$^*$ basis (in $D_{5h}$) and with Q-Chem\cite{qchem4} using a STO-3G basis, respectively. 

In table \ref{tab:illustrative_calculations} the number of CVs and their respective compression rate relative to the used basis set is listed together with the total wall-clock timings. Overall, the calculations finish in between two hours and two days, except for the largest system HBC calculated with an unc-cc-pVTZ basis with 2052 basis functions which took approximately four days. For all systems the numerical rank of the decomposition is just a fraction of the rank of a full decomposition and the compression rates are very large. For the buckminsterfullerene molecule with $\delta=5$ a compression rate of up to 277 is found which almost doubles the rates in comparison to a similar but field-free study performed in reference \onlinecite{Burger.2021}. This should not be mistaken to mean that the ff-CD is more efficient than its field-free counterpart. Instead, the reason for the discrepancy is because the full rank of the field-free CD is simply lower (see section \ref{sec:theory}). The high compression rate shows that the CD offers significant computational advantages compared to standard ff calculations.  

When loosening the Cholesky parameter to $\delta=4$, compression rates of up to 339 are observed. The computational savings are large enough to be in the order of a reduction of the cardinal number of the basis set, e.g., buckminsterfullerene calculated using a unc-dzp basis and $\delta=5$ is comparable in compuational time to a calculation using a unc-tzp basis and a threshold of $\delta=4$. While for $\delta=4$ the error in total energies is in the mHartree region, the error might be smaller for excitation energies: Reference \onlinecite{Krylov.2013} suggest that due to a systematic error cancellation, a parameter of $\delta=2-3$ together with an underlying HF calculation using the full ERI yields a sufficient accuracy. This will likely also be the case of excitation energies computed in a magnetic field but this hypothesis will need to be tested in future developments.

\begin{table*}[t]
\centering
\caption{Computational details for ff-CD-MP2 calculations on the systems corannulene, hexabenzocoronene (HBC), buckminsterfullerene, and retinal (see figure \ref{fig:Molecules}). The number of basis functions $N$, the Cholesky parameter $\delta$, the number of CVs $N_{\text{CH}}$, the compression rate, and the total wall time $t_{\text{tot}}$ are reported. All calculations were performed on 12 CPUs with the listed basis sets.}
\label{tab:illustrative_calculations}
\begin{ruledtabular}
\begin{tabular}{@{}llcccccr@{}}
Molecule & Basis & $N$ &Symmetry& $\delta$  & $N_{\text{CH}}$ & Compression rate & $t_{\text{tot}}$ [h:m:s] \\\midrule
\multirow{4}{*}{Corannulen} & \multirow{2}{*}{unc-cc-pVDZ} & \multirow{2}{*}{590} & \multirow{2}{*}{$C_{1}$} & 5 & 3835 & 90.77 & 1:59:53 \\
                     &                              &                          &                      & 4        & 2931              & 118.76      & 1:20:03   \\\cmidrule(l){2-8}
                     & \multirow{2}{*}{unc-cc-pVTZ} & \multirow{2}{*}{1000}    & \multirow{2}{*}{$C_{1}$} & 5        & 7916              & 126.33      & 14:51:43  \\
                     &                              &                          &                      & 4        & 6293              & 158.91      & 10:58:52  \\\midrule
\multirow{4}{*}{HBC} & \multirow{2}{*}{unc-cc-pVDZ} & \multirow{2}{*}{1218}    & $C_{1}$                  & 5        & 7985              & 185.79      & 15:49:25  \\
                     &                              &                          & $C_{2h}$                   & 5        & 8766              & 169.24      & 7:44:26   \\\cmidrule(l){2-8}
                     & \multirow{2}{*}{unc-cc-pVTZ} & \multirow{2}{*}{2052}    & $C_{1}$                  & 5        & 16372             & 257.19      & 111:28:26 \\
                     &                              &                          & $C_{2h}$                   & 5        & 17683             & 238.12      & 97:14:34 \\\midrule
\multirow{4}{*}{Fullerene}  & \multirow{2}{*}{unc-dzp}     & \multirow{2}{*}{1500}    & \multirow{2}{*}{$C_1$} & 5        & 9997              & 225.07      & 30:55:22  \\
                            &                              &                          &                      & 4        & 7792              & 288.76      & 20:00:35  \\\cmidrule(l){2-8}
                            & \multirow{2}{*}{unc-tzp}     & \multirow{2}{*}{1740}    & \multirow{2}{*}{$C_1$} & 5        & 10898             & 277.81      & 44:43:57  \\
                            &                              &                          &                      & 4        & 8918              & 339.49      & 31:18:59  \\\midrule
\multirow{4}{*}{Retinal}    & \multirow{2}{*}{unc-cc-pVDZ} & \multirow{2}{*}{742}     & \multirow{2}{*}{$C_1$} & 5        & 4739              & 116.18      & 2:09:15   \\
                            &                              &                          &                      & 4        & 3631              & 151.63      & 1:27:05   \\\cmidrule(l){2-8}
                            & \multirow{2}{*}{unc-cc-pVTZ} & \multirow{2}{*}{1330}    & \multirow{2}{*}{$C_1$} & 5        & 10467             & 169         & 41:33:25  \\
                            &                              &                          &                      & 4        & 8132              & 217.52      & 16:58:34  \\
\end{tabular}
\end{ruledtabular}
\end{table*}

A detailed report of the timings for computations on HBC is found in table \ref{tab:detailedtimings}. As seen by comparing serial with parallel calculations, the build of the CVs and following computational steps 
are parallelized. 
We note that parallelization is crucial in order to reduce the overall computational cost. 
So far, we only used threaded complex matrix-matrix multiplication routines (ZGEMM) from an appropriate BLAS library without any further attempts on efficient parallelization. 
The integral evaluation and the I/O for the CVs is currently only implemented in serial. 
Accordingly, the overall speed-up due to parallelization is not yet particularly convincing. 
In particular, it is observed that calculations that employ point-group symmetry gain very little speed-up through parallelization. For example, while without symmetry the speed-up of using 1 vs. 12 CPUs is 2.5, the serial implementation using symmetry is actually faster by a factor of 4.3. 
Performing the same calculation in parallel though only gives a factor of 5.0. 
Obviously, the parallelization might be improved drastically outside of the BLAS routines which is however outside of the scope of the present paper. 

While for small systems the integral evaluation is the bottleneck of the calculation, for larger systems this shifts to the SCF iterations. This is also partly due to the fact that for such large systems and basis sets convergence is more difficult requiring need methods like DIIS and damping as well as more iterations in general. 
It will hence be useful to adapt second-order methods for finite-field calculations to accelerate and ensure SCF convergence in the future.\cite{QCSCF1,QCSCF2,Lipparini.2021} 
\begin{table}[htbp]
\centering
\caption{Detailed wall clock timings in minutes for a ff-CD-MP2/unc-cc-pVDZ calculation (N=1218) on hexabenzocoronene in a homogeneous magnetic field perpendicular to the molecular plane of the strength $0.1\B$ and a Cholesky parameter of $\delta=5$.}
\label{tab:detailed_timings}
\begin{ruledtabular}
\begin{tabular}{@{}ccccc@{}}
                                                  & \multicolumn{2}{c}{$C_1$} & \multicolumn{2}{c}{$C_{2h}$} \\ 
                                    \cmidrule(lr){2-3}
                                     \cmidrule(lr){4-5}
time in min                                            & serial   & parallel   & serial   & parallel   \\[2pt]\midrule
$t_{\text{int}}$                               & 153.4     & 152.4           & 191.3     & 191.2          \\[2pt]
$t_{\text{chol}}$\protect\footnote{For the timings of $t_{\text{chol}}$ only the build of the CVs as in equation \eqref{eq:cholvec} is considered which is the most expensive step of the Choleksy procedure scaling $\mathcal{O}(N^2N_{\text{CH}}^2)$.}  & 709.6     & 217.3           & 186.8     & 185.1          \\[2pt]
$t_{\text{i/o}}$    & 2 x 1.5       & 2 x 1.9             & 2 x 0.6       & 2 x 0.6            \\[2pt]
$t_{\text{HF}}$\protect\footnote{Timings 'per iteration' are given in parentheses.}        & 877.4 (15.7)     & 408.5 (8.7)           & 55.4 (2.6)      & 50.9 (2.4)           \\[2pt]
$t_{\text{ao2mo}}$     & 55.9      & 15.2            & 6.1       & 1.8            \\[2pt]
$t_{\text{MP2}}$       & 529.5     & 144.8           & 102.9     & 29.4           \\[2pt]
$t_{\text{tot}}$    & 2336.4    & 949.4           & 548.6     & 464.4          \\ 
\end{tabular}
\end{ruledtabular}
\end{table}
\section{Conclusion}\label{sec:conclusion}

In this work, we report on a CD scheme for the calculation of large molecules in finite magnetic fields at the MP2 level of theory. 
The scheme expands the range of applicability to systems with more than 2000 basis functions. 
A complex-valued implementation that exploits point-group symmetry 
was presented which performs a CD of the two-electron integrals over GIAOs and uses the CVs in subsequent finite-field HF and MP2 computations. 
We showed that the accuracy of the energy scales with the Cholesky parameter $\delta$ which results in a rigorous error control via a user-defined Cholesky parameter. 
We also showed that the compression rate which is achieved by the CD depends strongly on the strength and orientation of the magnetic field. 
In particular, the number of required CVs increases with the strength of the magnetic field while at the same time still yielding very high compression rates. Furthermore, for linear molecules the number of CVs is maximal in the perpendicular orientation with respect to the  magnetic field. 
Due to the fact that in the general case the ERIs only exhibit fourfold permutational symmetry the possible savings in terms of compression rate are particularly high in finite-field calculations. 
Noting that the CD retrieves the necessary CVs in a black-box manner with a pre-definable accuracy makes it very well suited for the computations in varying magnetic-field strengths and orientations. 
This is a clear advantage over the use of RI methods for which auxiliary basis sets need to be employed which lack rigorous error control. 
So far, only magnetic-field independent auxiliary basis sets have been used in ff calculations. 
Because of the well-known connection between CD and RI, ff-CD results might be employed in the generation of field-dependent auxiliary basis sets for the use within RI calculations.
The fact that the number of CVs changes so drastically with field-strength and orientation indicates that such a task, i.e., generating auxiliary sets with a solid and reliable accuracy for different field strengths and orientations, may be challenging.
Quantum-chemical calculations in strong magnetic fields typically explore unknown terrains, making the reliability of the predictions even more important.
This work also constitutes a first step towards a highly accurate treatment of larger systems in strong magnetic fields. 
It enables studies on -for example- the paramagnetic-to-diamagnetic transition of large paramagnetic closed-shell molecules which may occur at much lower and hence experimentally accessible magnetic-field strengths than what is predicted for small systems.\cite{TellgrenPhys.Chem.Chem.Phys2009} 
In addition, the use of CD allows moving to larger basis sets and hence towards higher accuracy which is particularly important for spectroscopic predictions. Following the work of Ref. \onlinecite{Krylov.2013}, the CD can also successfully be applied in more sophisticated post-HF methods.
As the basis-set error is often the limiting factor in terms of accuracy in high-level ff calculations, the combination of CD with ff ground-state and excited-state coupled-cluster methods$^{\cite{Stopkowicz.2015,Hampe.J.Chem.Phys.2017,Stopkowicz.2018,Hampe.J.Chem.TheoryComput.2019,HampePhys.Chem.Chem.Phys.2020}}$  will be a beneficial future direction of development. 

\begin{acknowledgments}
The authors thank Jürgen Gauss and Filippo Lipparini for helpful discussions. This work has been supported by DFG grant number STO 1239/1-1

\end{acknowledgments}

\appendix

\section{data availability statement}

The data that supports the findings of this study are available within the article and its {supplementary material}.


\providecommand{\noopsort}[1]{}\providecommand{\singleletter}[1]{#1}%

\end{document}